\documentclass[notoc,letterpaper]{JHEP}
\usepackage{epsfig}
\usepackage{axodraw}

\def\PL#1#2#3{{\it Phys. Lett. }{\bf B#1~}(#2)~#3}
\def\PRL#1#2#3{{\it Phys. Rev. Lett. }{\bf #1~}(#2)~#3}
\def\PREP#1#2#3{{\it Phys. Rep. }{\bf #1~}(#2)~#3}
\def\PRD#1#2#3{{\it Phys. Rev. }{\bf D#1~}(#2)~#3}
\def\NPB#1#2#3{{\it Nucl. Phys. }{\bf B#1~}(#2)~#3}
\def\JHEP#1#2#3{JHEP #1:#3~(#2)}
\def\xxx#1{\texttt{[#1]}}

\def\mg{{M_{GUT}}}

\def\qslash{\not{\hbox{\kern-2pt $q$}}}
\def\delslash{\not{\hbox{\kern-2pt $\partial$}}}
\def\Dslash{\not{\hbox{\kern-2pt $D$}}}

\def\del{\partial}
\def\beq{\begin{equation}}
\def\eeq{\end{equation}}
\def\eeq{\end{equation}}
\def\bea{\begin{eqnarray}}
\def\eea{\end{eqnarray}}
\def\bq{\begin{quote}}
\def\eq{\end{quote}}
\def\lesssim{\mathrel{\mathpalette\vereq<}}
\def\gtrsim{\mathrel{\mathpalette\vereq>}}
\def\lsim{\mathrel{\lesssim}}
\def\gsim{\mathrel{\gtrsim}}

\makeatletter
\def\vereq#1#2{\lower3pt\vbox{\baselineskip1.5pt \lineskip1.5pt
\ialign{$\m@th#1\hfill##\hfil$\crcr#2\crcr\sim\crcr}}}
\makeatother

\title{\center{Gaugino-Assisted Anomaly Mediation}}
\author{David~Elazzar~Kaplan \\ High Energy Physics Division,
Argonne National Laboratory, Argonne, IL 60439 \\ Enrico Fermi Institute,
Univ. of Chicago, 5640 Ellis Avenue, Chicago, IL 60637 \\
\email{dkaplan@theory.uchicago.edu}}

\author{Graham~D.~Kribs \\ Department of Physics, Carnegie Mellon University,
Pittsburgh, PA 15213-3890 \\ 
Department of Physics, University of Wisconsin, 1150 University Avenue, 
Madison, WI 53706.\footnote{Present address.} \\ 
\email{kribs@pheno.physics.wisc.edu}}

\preprint{\hepph{0009195} \\ ANL-HEP-PR-00-090, EFI 00-27 \\ MADPH-00-1193}
\abstract{We present a model of supersymmetry breaking mediated through
a small extra dimension.  Standard model matter multiplets and a 
supersymmetry-breaking (or ``hidden'') sector are confined to opposite
four-dimensional boundaries while gauge multiplets live in the bulk.  
The hidden sector does not contain a singlet and the dominant 
contribution to gaugino masses is via anomaly-mediated supersymmetry
breaking.  Scalar masses get contributions from both
anomaly mediation and a tiny hard breaking of supersymmetry by
operators on the hidden-sector boundary.  These operators contribute to 
scalar masses at one loop and in most of parameter space, their
contribution dominates.  Thus it is easy to make all squared 
scalar masses positive.  As no additional fields or symmetries 
are required below the Planck scale, we consider this the simplest 
working model of anomaly mediation.  The gaugino spectrum is left 
untouched and the phenomenology of the model is roughly similar to 
anomaly mediated supersymmetry breaking with a universal scalar mass 
added.  We identify the main differences in the spectrum between this 
model and other approaches.  We also discuss mechanisms for generating 
the $\mu$ term and constraints on additional bulk fields.}


\begin{document}

\section{Introduction}
\label{intro}
If supersymmetric partners of standard model particles are discovered
by the collider experiments of the coming decade, one of the first theoretical
questions that must be answered is ``how did supersymmetry break?''.  Softly
broken supersymmetry as an effective low energy theory contains over one
hundred parameters, but we expect there is an organizing principle 
that determines most of these parameters.  This involves a mechanism 
to break supersymmetry and a means to communicate, or mediate, 
supersymmetry breaking to the superpartners of the Standard Model (SM).  
It behooves us to search for simple and compelling models that generate
soft masses of order the weak scale in a predictive and experimentally
allowed fashion.

Viable models are those which do not have contributions to processes beyond
their experimental bounds.  Processes which include, for example, flavor
changing neutral currents (FCNC), CP violation or lepton flavor violation,
are suppressed in the standard model, and thus whatever generates the soft 
parameters must also sufficiently suppress flavor and 
CP violation\cite{align,FCNC}.  The former is naturally suppressed in models
of gauge-mediated supersymmetry breaking (GMSB), where soft terms come from
loop contributions involving standard model gauge interactions \cite{GMSB}.
Since the gauge interactions are flavor blind, the soft masses are as well.

Another candidate mechanism for mediating supersymmetry breaking is via
the superconformal anomaly \cite{RS,GLMR,BMP}.  In the absence of
pure singlets in the supersymmetry breaking sector, gaugino masses 
are generated at the one-loop level \cite{GLMR}.  If in addition the standard 
model fields are confined to a four-dimensional boundary in a
higher-dimensional space and supersymmetry breaking occurs on a different
boundary (the hidden sector) spatially separated from the standard model, 
then the dominant contribution to scalar masses come from anomaly mediation and 
are of the same order as the gaugino masses \cite{RS}.  Large contributions
to FCNC processes are avoided because the scalar masses, in particular 
the first and second generation, are dominated by contributions 
proportional to the beta functions of gauge couplings.  Similarly, 
CP violation can also be suppressed if all weak scale masses are
generated by a single hidden sector parameter.


However, pure anomaly mediation predicts tachyonic sleptons.  
The squared scalar mass contributions are proportional to beta functions of 
gauge couplings, with non-asymptotically free gauge groups giving them
negative contributions.  In the minimal supersymmetric standard
model (MSSM) the SU(2)$\times$U(1) gauge groups are non-asymptotically 
free, and thus the squared masses of sleptons (in particular, those of 
the first two generations) have overwhelmingly dominant contributions 
which are negative.  Clearly, additional model building is required.  
The task is nontrivial since the form of the soft parameters are 
renormalization group invariant, depending only on the infrared values 
of the beta functions \cite{RS,GLMR}.

Nevertheless, interesting solutions which avoid these difficulties
have been suggested.  Pomarol and Rattazzi proposed adding a light singlet 
to the visible sector which obtains a large vacuum expectation value (vev) 
for its scalar component, and then used this scale to generate 
a threshold that does not decouple \cite{PR}.   Thus a spectrum very
different from that of the original anomaly-mediated models is produced 
\cite{PRpheno} and the slepton mass problem can be solved.  This idea has 
also been applied to a GUT threshold in Ref.~\cite{CLPSS}.
Katz, Shadmi, and Shirman showed that threshold effects do not
decouple at higher orders in the supersymmetry breaking order parameter, 
but this effect is only significant for thresholds that are not too far 
above the weak scale \cite{KSS}.
Another possibility is to alter the infrared structure of the
MSSM directly, by adding new matter charged under the standard model
and new Yukawa couplings \cite{CLMP} or an additional U(1) gauge group 
\cite{PR,KSS,JackJonesFI,CHK}.
Finally, one can add additional 
non-MSSM fields to the bulk to generate new contributions to soft masses
\cite{bulkU1,KKS,CLNP}, though in most cases, these contributions dominate
those of anomaly mediation.
In any case, the additional structure needed to solve the slepton
mass problem invariably affects much more of the sparticle spectrum 
than just the slepton masses, and so the phenomenology need not
have any resemblance to ``pure'' anomaly mediation.

In this article, we present a simple model of supersymmetry breaking
mediated by both supergravity and MSSM gauge multiplets
that live in the higher dimensional ``bulk''.  The hidden sector is 
spatially separated from the matter sector, and consequently 
contact interactions are forbidden by locality.  In this way, the model
is similar to gaugino-mediated supersymmetry breaking 
(\~{g}MSB) \cite{KKS,CLNP} as well as anomaly mediation \cite{RS, GLMR}.
However, unlike the original gaugino mediation proposals \cite{KKS,CLNP}, 
we assume there are no fundamental singlets in the hidden sector.  
This is well motivated as most known models of dynamical supersymmetry 
breaking lack singlets \cite{ADS,DM}, and was a major motivation that led 
to uncovering anomaly-induced gaugino masses \cite{GLMR}.
We will show that this implies the dominant contribution to gaugino 
masses is from anomaly mediation.  Scalar masses also receive their usual 
contribution from anomaly mediation, but in this model the gauge and 
gaugino fields can also communicate supersymmetry breaking to the matter 
sector.  We will show that the leading gaugino-mediated contribution 
comes from a set of higher dimensional operators which generically appear
on the hidden sector boundary.  These operators result in tiny corrections
to the wave function renormalization of gauginos different from gauge fields 
and thus are a hard breaking of supersymmetry.  
These operators contribute to squared scalar masses at one loop.
In doing a full five-dimensional calculation, we show that their contributions
are of the right size to solve the slepton mass problem 
of anomaly mediation.  Below the compactification scale, the model 
therefore has all the appearances of anomaly mediation with an additional 
contribution to scalar masses.  This is perhaps the most minimal
solution to the slepton mass problem in anomaly mediation.
Notice that because the interactions that give rise to the additional 
scalar mass contributions involve gauge and gaugino fields, they are 
automatically flavor-blind.  We need only assume that the extra 
dimension is large enough such that the exponentially suppressed wave-function 
overlap estimates of contact FCNC interactions are indeed suppressed without 
fine tuning (a compactification length one order of magnitude larger than 
the five-dimensional Planck length is sufficient).

The rest of this paper is organized as follows.  In Section \ref{amsb} we 
review the salient features of anomaly mediation including the predicted 
spectrum and its decoupling features.  In Section \ref{gaam-sec} 
we present our model.  We show that the dominant
contribution to gaugino masses is from anomaly mediation, while the
bulk gauge and gaugino fields give rise to additional contributions to 
scalar masses.  In Section \ref{pheno} we discuss phenomenology, including 
the list of input parameters and the generic spectrum.  
In Section \ref{bulk} we briefly discuss solving the $\mu$ problem
and the effects of adding bulk fields and in Section \ref{conclusion} we 
conclude.  An appendix is included in which we demonstrate how we take
into account the effects of GUT fields at the threshold.

\section{Anomaly mediated supersymmetry breaking}
\label{amsb}

In this section we review the contributions to soft terms from the 
superconformal anomaly \cite{RS,GLMR} as well as the construction
in extra dimensions which makes these contributions the dominant ones
\cite{RS}.  There are actually several distinct contributions to, 
for instance, gaugino masses from the anomalous symmetries in the 
supergravity Lagrangian \cite{BMP} (see also \cite{GN}).  These 
contributions reduce to that of just the superconformal anomaly 
\cite{RS,GLMR} when it is assumed that all vacuum expectation values 
in or coupled to the hidden sector are much smaller than the Planck mass, 
$M_{\rm Pl}$.  For the remainder of this paper, we shall make this 
assumption.

If singlets are absent from the hidden sector, the dominant contribution
to gaugino masses comes from anomaly mediation and is loop suppressed
with respect to the gravitino mass, $m_{3/2}$.  Scalar masses would in 
general be of order $m_{3/2}$ as usual in ``SUGRA'' scenarios, meaning
that not only does one still have the usual FCNC problem, but one must
also fine-tune the coefficients of at least some of the operators 
to be of order a loop factor \cite{GLMR}.  However, scalar masses 
can be suppressed by spatially separating the hidden sector from the 
MSSM in extra dimensions.  For our purposes, we imagine one extra 
flat spatial dimension compactified on $S^1/Z_2$.  The fifth coordinate 
$y$ runs from $-L$ to $L$ with $y\rightarrow -y$ identified.  
The four-dimensional hypersurfaces at $y=0$ and $L$ are where the 
MSSM and hidden sectors live respectively, while supergravity
lives in the bulk \cite{RS}.  The compact dimension could be stabilized
by the mechanisms in Ref. \cite{LS} or \cite{AHHSW} without greatly changing
the dynamics of supersymmetry breaking.

With an extra dimension of size $L$, the five-dimensional
Planck scale $M_*$ is related to the four-dimensional (reduced) Planck 
scale $M_{\rm Pl}$ by \cite{ADD}
\begin{equation}
M_*^3 L = M_{\rm Pl}^2.
\label{add}
\end{equation}
To solve the flavor problem, we must suppress flavor violating operators
such that the off-diagonal squared squark masses are at most of order $10^{-3}$
times the diagonal squared squark masses \cite{FCNC}.  Thus, operators like
\begin{equation}
\frac{1}{M_*^2}\int d^4\theta \Sigma^{\dagger}\Sigma Q_i^{\dagger} Q_j 
	\:\:\:\:\:\:(i\neq j)
\end{equation}
should conservatively have a coefficient less than 
of order $10^{-3}\times (m_{weak}^2/m_{3/2}^2) \sim 10^{-7}$.  
These operators 
could be generated by the exchange of particles of mass $M_*$ and thus 
are suppressed by the exponential suppression of the Yukawa 
potential $e^{-M_* L}$, requiring $M_*L\gsim 16$.

To calculate the soft masses, it is simplest to use the compensator 
formalism of supergravity \cite{crem}, which has been discussed 
in the original anomaly mediation papers \cite{RS, GLMR}.
In this approach, superconformal symmetry is made manifest by using
the conformal compensator $\Phi$ as a spurion for symmetry breaking,
where
\begin{equation}
\Phi=1+\theta^2 m_{3/2}.
\end{equation}
In this formalism, $\Phi$ is the only source of supersymmetry breaking
in the visible sector.  The coupling of the MSSM to supergravity can
be written
\begin{eqnarray}
\cal{L} &=& \int d^4 \theta \Phi^\dag \Phi f(Q^\dag, e^{-V} Q) 
+ \int d^2 \theta \Phi^3 \left( W(Q) + \frac{1}{g^2} 
  \mathcal{W}_\alpha \mathcal{W}^\alpha \right) + h.c. + \ldots
\end{eqnarray}
where we have written the K\"ahler potential, superpotential, and
gauge superfield terms in flat space.  Each superfield can be rescaled by 
$\Phi^{-n}$ where $n$ is the canonical dimension of the field.  Thus, in the MSSM 
without a $\mu$ term, supersymmetry breaking does not appear at tree level.
However, all dimensionful parameters will come with the appropriate power
of $\Phi$.  The key observation is that conformal symmetry is broken at the 
loop level by any regulator of the theory, and that the cutoff scale must 
also appear with the compensator.  

Thus, we expect soft masses to appear at loop level.  These masses can be
obtained via explicit calculation with a regulator \cite{RS,GLMR} or using
the method of `analytical continuation into superspace' \cite{GR,GLMR,PR}.  
In any case, one finds \cite{RS,GLMR}
\begin{eqnarray}
M_{a} & = & \frac{\beta_{g_a}}{g_a} m_{3/2} \nonumber\\ 
\tilde{m}^2 _i & = &-\frac{1}{4}\left(\frac{\del \gamma_i}{\del g}
	\beta_g + \frac{\del\gamma_i}{\del Y} \beta_Y \right)
	m_{3/2}^2 \label{AMsoft}\\
A_Y & = & - \frac{\beta_Y}{Y} m_{3/2}.\nonumber
\end{eqnarray}
There are three important aspects of this spectrum to notice.  First, 
flavor violation is proportional to the Yukawa couplings.  Second, because
the beta functions for SU(2) and U(1) are both positive, the squared masses
of at least the first two generations of sleptons are negative.
Third, the forms of these soft masses are renormalization group 
invariant to all orders \cite{GLMR, JackJonesRGinvariant}.
In fact, supersymmetric mass thresholds that are significantly above 
the weak scale completely decouple \cite{GLMR,PR,KSS}.  This can be seen
as arising from a cancellation between the anomaly-induced supersymmetry 
breaking contributions with threshold contributions.
Hence, the results in Eqs.~(\ref{AMsoft}) are valid at the weak scale.
This presents a significant obstacle in solving the slepton mass problem.

The simplest approach to dealing with the slepton mass problem
is to simply add an additional universal scalar mass squared $m_0^2$ to the
spectrum \cite{GGW, FManom}.  This is really just a phenomenological 
fix, however, and levies the explanation of why FCNC are suppressed 
into the postulate that $m_0^2$ is flavor-diagonal.  Nevertheless, 
this does allow an analysis of the phenomenology of gaugino masses 
proportional to the gauge beta functions.

As discussed in the introduction, there have been several groups that 
have specific proposals for solving the slepton mass problem.  
In all cases, the goal is to circumvent the decoupling phenomenon.  
All of these proposals require the addition of additional matter and/or 
gauge groups to the MSSM and all but those in \cite{JackJonesFI, CHK}
significantly change the gaugino spectrum\footnote{Adding weak-scale 
Fayet-Illiopoulos terms for hypercharge and an additional U(1) in the
MSSM does not affect the gaugino spectrum \cite{JackJonesFI,CHK} and thus
a model in which these terms are generated dynamically at the correct
size would be quite interesting.}.

Thus, anomaly mediation by itself is a fascinating but phenomenologically
unacceptable means to communicate supersymmetry breaking.  Any model
in which the AMSB contributions are important (or dominant) must
be supplemented by a true solution to the slepton mass problem, 
and this generally requires rather complicated additional structure
that need not leave the spectrum looking anything like simply
adding a universal scalar mass.  


\section{Gaugino assisted anomaly mediation}
\label{gaam-sec}

Consider taking the model of anomaly mediation in its original form 
\cite{RS}, but place the MSSM gauge and gaugino fields in the bulk.  
If singlets in the hidden sector exist, the dominant contribution to 
soft masses comes from gaugino mediation \cite{KKS, CLNP}, since 
the anomaly contributions are loop suppressed.  Now, consider 
the scenario in which no singlets exist on the hidden sector boundary.
Singlets are absent in many models of dynamical supersymmetry breaking,
and so this can be thought of as a well-motivated ``special case'' 
of gaugino mediation.  We shall show below that, in this case,
anomaly mediation plays a much more important role, but that
the presence of gauge and gaugino fields in the bulk results in 
additional contributions to scalar masses that are naturally of the 
same size as their anomaly-mediated counterparts.  The result is an 
explicit model in which anomaly-mediated contributions are large
(and dominant for the gaugino masses), while the slepton masses
can easily be positive.

There are a few subtleties of placing gauge and gaugino fields 
in $4 + 1$ dimensions.  The vector multiplet has twice as many physical
degrees of freedom.  The gauge field $A^{\mu}$ has a fifth component $A^5$, 
there are two $2$-component gaugino spinors, $\lambda^1$ and $\lambda^2$, 
and there is a real scalar field $\Phi$, all in the adjoint representation.  
We can selectively decouple the effects of some of these fields by 
making them odd under the $Z_2$ part of the compactification, 
explicitly breaking half of the ($N=2$ in four dimensions) supersymmetries
at the boundary.  Then, the contributions of these fields to what 
follows are negligible.  For a detailed prescription and analysis, 
see Ref.~\cite{MP}.

Putting gauge fields in the bulk allows for local operators
combining the gauge multiplets with fields of the hidden sector.  
For instance, if the hidden sector contains a singlet $S$, 
gaugino masses are generated by the operator
\begin{equation}
\int d^2 \theta \frac{S}{M_*^2} W^{\alpha} W_{\alpha} \delta(y-L)
\label{gaugino}
\end{equation}
which gives $M_{1/2}= F_S/(M_*^2 L)= (F_S/M_{\rm Pl})\times (M_* L)^{-1/2}$.
This can be compared with the anomaly-mediated contribution for a 
gravitino mass $m_{3/2} \equiv F/M_{\rm Pl}$ where we associate
the fundamental $F$ term with $F_S$, resulting in 
$M_{1/2}^{\rm AM} = (16 \pi^2)^{-1} F_S/M_{\rm Pl}$.
For $M_*L \ll 10^4$ (which is always satisfied if $M_*$ is the scale
where either the gauge or gravitational coupling gets strong), the
gaugino-mediated contribution dominates, and the result is gaugino 
mediated supersymmetry breaking \cite{KKS,CLNP}.
Scalar masses are then generated at the weak scale via the renormalization
group and the spectrum looks similar to that of no-scale models 
\cite{noscale}, with differences depending on the compactification scale 
and how the $\mu$ term is generated \cite{KKS,CLNP,SS1,SS2,KT}.

If we now suppose that there are no singlets in the hidden sector, 
the operator (\ref{gaugino}) is absent.  In the K\"ahler potential, 
gaugino masses can be generated but they are highly suppressed.
The leading higher-dimension operator with the least powers of $M_*$ is
\begin{eqnarray}
\int d^4 \theta && \frac{\Sigma^\dag \Sigma}{M_*^4} 
W^\alpha W_\alpha \delta(y - L) \; .
\end{eqnarray}
where $\Sigma$ is a hidden sector field charged under some hidden
sector group(s) with the largest auxiliary 
component vev $F_{\Sigma}$.  However, it is obvious that this just gives
\begin{eqnarray}
\frac{F_\Sigma^\dag}{M_*^2} \int d^2 \theta && \frac{\Sigma}{M_*^2} 
W^\alpha W_\alpha \delta(y - L)
\end{eqnarray}
and so takes the same form as (\ref{gaugino}) except that
there is the additional suppression of $F/M_*^2$.  Hence, the contribution
arising from this operator is of order $F^2/(M_*^4 L)$, which
is negligible compared with the anomaly-mediated contribution.

However, even though (\ref{gaugino}) is not allowed, there are a number 
of operators expected to appear on the hidden sector boundary which can 
contribute to the spectrum.  Some allowed operators are 
\begin{eqnarray}
\int d^4\theta &&\frac{\Sigma^{\dagger}D^2 \Sigma}{M_*^5} 
	W^{\alpha} W_{\alpha} \delta(y-L), \label{opsa} \\
\int d^4\theta &&\frac{\Sigma^{\dagger}\Sigma}{M_*^5} 
	W^{\alpha}D^2 W_{\alpha} \delta(y-L), \label{opsb} \\
\int d^4\theta &&\frac{\Sigma^{\dagger}\Sigma}{M_*^5} 
	D_{\beta}W^{\alpha}D^{\beta} W_{\alpha} \delta(y-L), \label{opsc} \\
\int d^4\theta &&\frac{\Sigma^{\dagger}D_{\beta}\Sigma}{M_*^5} 
	W^{\alpha}D^{\beta} W_{\alpha} \delta(y-L), \label{opsd} \\
\int d^4\theta&&\frac{\Sigma^{\dagger}\Sigma}{M_*^5} 
	D^{\beta} W^{\alpha}D^{\alpha} W_{\beta} \delta(y-L), \label{opse} \\
\mathrm{etc...}&&\phantom{\frac{\Sigma^{\dagger}\Sigma}{M_*^5}}  \label{opsf}
\end{eqnarray}
where $D_{\alpha}$ is the superspace derivative and the ellipsis denotes
additional terms similar to those above.  When $\Sigma$ is replaced 
with its $F$ component, these operators become tiny corrections to 
the kinetic terms of the gauge multiplet.  For instance, Eq.~(\ref{opsa})
becomes
\begin{eqnarray}
\int & d^4\theta & \frac{\Sigma^{\dagger}D^2 \Sigma}{M_*^5}
        W^{\alpha} W_{\alpha} \delta(y-L)\nonumber\\
	&\rightarrow&
	\int d^2\theta \frac{F_{\Sigma}^{\dagger} F_{\Sigma}}{M_*^4 (M_* L)}
	W^{0\alpha} W_{\alpha}^0\nonumber\\
	&\rightarrow&
	\frac{F_{\Sigma}^{\dagger} F_{\Sigma}}{M_*^4 (M_* L)}
	\left[ -2i\lambda \sigma^{\mu} D_{\mu} \overline{\lambda} 
	-\frac{1}{2} F^{\mu\nu}F_{\mu\nu} + D^2 
	+ \frac{i}{4}F^{\mu\nu}\widetilde{F}_{\mu\nu}\right],
\end{eqnarray}
where the five-dimensional supersymmetric field strength $W^{\alpha}$ has 
been replaced by its (rescaled) zero mode $W^{0\alpha}$, and $F^{\mu\nu}$,
$\lambda$ and $D$ are the four-dimensional gauge field strength, gaugino
and auxiliary components respectively.  Thus the effect 
of this operator is to simply rescale the gauge coupling by a tiny amount.

The other operators in Eqs.~(\ref{opsb})--(\ref{opsf}), however, 
have non-negligible effects.  For example, the operator Eq.~(\ref{opsb}) 
becomes \cite{KKS}
\begin{equation}
\int d^4\theta \frac{\Sigma^{\dagger}\Sigma}{M_*^5}
        W^{\alpha}D^2 W_{\alpha} \delta(y-L) \rightarrow
	\frac{\left| F_{\Sigma}\right|^2 }{M_*^4}\frac{1}{(M_* L)}
	\lambda \sigma^{\mu} D_{\mu} \overline{\lambda},
\label{gauginokinetic}
\end{equation}
where $\lambda$ is the four-dimensional zero-mode of the gaugino.  Note
that this operator contributes \emph{only} to the gaugino kinetic term.
This corresponds to a wave-function renormalization of the gaugino
slightly different from that of the gauge field and auxiliary components
and thus leads to a (albeit tiny) hard breaking of supersymmetry.

First, let us make a rough estimate of the effects of this hard breaking
on the spectrum.
To do this, we rescale the gaugino wave function such that its kinetic 
term is canonical at the compactification scale $L^{-1}$.  
Such a rescaling changes (for example) the gaugino-quark-squark 
coupling $g_{\tilde{g}q\tilde{q}}$ relative to the gauge coupling $g$ 
as\footnote{A shift of the gaugino coupling relative to the 
gauge coupling also arises from superoblique corrections \cite{superoblique},
but this is completely distinct from the operators we are 
adding.}
\begin{equation}
g_{\tilde{g}q\tilde{q}}\rightarrow g'_{\tilde{g}q\tilde{q}} \simeq
	\left(1 + \frac{F_\Sigma^2}{M_*^5 L}\right)g .
\end{equation}
Now the one-loop calculation of the contribution to squared scalar masses 
appears quadratically divergent.  Taking $\Lambda = L^{-1}$ to be 
the cutoff, we can approximate this contribution as
\begin{eqnarray}
\delta m_{\tilde f}^2 &\sim& \frac{g^2}{16 \pi^2} 
	\frac{\left| F_{\Sigma} \right|^2}{M_*^5 L} \Lambda^2 \nonumber\\
&\sim& \frac{g^2}{16 \pi^2}
        \frac{\left| F_{\Sigma} \right|^2}{M_{\rm Pl}^2}\frac{1}{(M_* L)^2} .
\end{eqnarray}
From this estimate it is clear that there is some range of compactification
scales $L^{-1}$ for which this contribution is as or more important as
the one from anomaly mediation.  Notice that these contributions ought
to be present in any effective theory.  In particular, in
ordinary supergravity models we can write the operator
\begin{eqnarray}
\int d^4 \theta \frac{\Sigma^\dag \Sigma}{M_{\mathrm{Pl}}^4} 
W^\alpha D^2 W_\alpha 
\end{eqnarray}
(where now the field strength superfield $W^\alpha$ is purely 
four-dimensional) which also leads to contributions to scalar masses of 
order $F^2/M_*^2$ times a loop factor.

\FIGURE[t]{
\begin{picture}(420,120)(-30,0)
%
%
   \Text(  25,   5 )[c]{(a)}
   \Line(   1,  90 )(   1,  30 )
   \Line(  20, 110 )(  20,  50 )
   \Line(   1,  90 )(  20, 110 )
   \Line(   1,  30 )(  20,  50 )
   \Line(  36,  90 )(  36,  30 )
   \Line(  55, 110 )(  55,  50 )
   \Line(  36,  90 )(  55, 110 )
   \Line(  36,  30 )(  55,  50 )
   \DashLine(    5,  45 )(   9,  62 ){2}
   \Line(        9,  62 )(  12,  78 )
   \DashLine(   12,  78 )(  15,  95 ){2}
   \CArc(       22,  51 )(  30, 40, 109 )
   \PhotonArc(  22,  51 )(  30, 40, 109 ){2}{8}
   \CArc(       20,  88 )(  30, 247, 320 )
   \PhotonArc(  20,  88 )(  30, 247, 320 ){2}{8}
   \put(   46, 70 ){\circle*{6}}
%
%
   \Text( 125,   5 )[c]{(b)}
   \Line( 101,  90 )( 101,  30 )
   \Line( 120, 110 )( 120,  50 )
   \Line( 101,  90 )( 120, 110 )
   \Line( 101,  30 )( 120,  50 )
   \Line( 136,  90 )( 136,  30 )
   \Line( 155, 110 )( 155,  50 )
   \Line( 136,  90 )( 155, 110 )
   \Line( 136,  30 )( 155,  50 )
   \DashLine(  105,  45 )(  115,  95 ){2}
   \PhotonArc( 128,  63 )(  20, 25, 155 ){2}{8}
   \PhotonArc( 128,  80 )(  20, 205, 335 ){2}{8}
   \put(  146, 70 ){\circle*{6}}
%
%
   \Text( 225,   5 )[c]{(c)}
   \Line( 201,  90 )( 201,  30 )
   \Line( 220, 110 )( 220,  50 )
   \Line( 201,  90 )( 220, 110 )
   \Line( 201,  30 )( 220,  50 )
   \Line( 236,  90 )( 236,  30 )
   \Line( 255, 110 )( 255,  50 )
   \Line( 236,  90 )( 255, 110 )
   \Line( 236,  30 )( 255,  50 )
   \DashLine(  205,  45 )( 209, 62 ){2}
   \DashLine(  209,  62 )( 212, 78 ){2}
   \DashLine(  212,  78 )( 215, 95 ){2}
   \PhotonArc( 222,  51 )(  30, 40, 109 ){2}{8}
   \PhotonArc( 220,  88 )(  30, 247, 320 ){2}{8}
   \put(  246, 70 ){\circle*{6}}
%
%
   \Text( 325,   5 )[c]{(d)}
   \Line( 301,  90 )( 301,  30 )
   \Line( 320, 110 )( 320,  50 )
   \Line( 301,  90 )( 320, 110 )
   \Line( 301,  30 )( 320,  50 )
   \Line( 336,  90 )( 336,  30 )
   \Line( 355, 110 )( 355,  50 )
   \Line( 336,  90 )( 355, 110 )
   \Line( 336,  30 )( 355,  50 )
   \DashLine(  305,  45 )( 309,  62 ){2}
   \DashLine(  309,  62 )( 312,  78 ){2}
   \DashLine(  312,  78 )( 315,  95 ){2}
   \DashCArc(  322,  51 )(  30, 40, 109 ){2}
   \DashCArc(  320,  88 )(  30, 247, 320 ){2}
   \put(  346, 70 ){\circle*{6}}
\end{picture}
\caption{One loop contributions to scalar masses.  The dot
represents the operator insertion in the five-dimensional propagators.
The five-dimensional propagators in the loops are (a) gauginos, 
(b) \& (c) gauge bosons, and (d) the real scalar adjoint $\Phi$.}
\label{loop}}

Now we examine this new scalar mass contribution more carefully.  
The full one-loop calculation is extremely well 
approximated by a five-dimensional loop diagram with a single 
insertion of the operator (\ref{gauginokinetic}) on the gaugino 
propagator (see Fig.~\ref{loop}a).  Using the mixed 
momentum/position space propagators $\mathcal{P}$ as defined in the
Appendix (see also Ref.~\cite{MP,AHGS,KKS}), we have
\begin{equation}
\tilde{m}^2_i = \frac{2 g_{(5)}^2 C(i)}{16 \pi^2} \int d^4 q \,\, 
\mathrm{tr}\left[ \frac{-1}{\qslash} \mathcal{P}(q;0,L) 
\frac{\qslash F_\Sigma^2}{M_*^5} \mathcal{P}(q;L,0) \right]
\end{equation}
where $g_{(5)}$ is the five-dimensional gauge coupling, $C(i)$ is 
the quadratic Casimir for the $i$ matter scalar representation, 
and the integral is over all (Euclidean) four-momenta.  The 
integrand is highly peaked around the compactification scale so 
when computing the spectrum, it is reasonable to begin 
renormalization group evolution at $L^{-1}$.  
The integral over momenta is finite, due to physical point-splitting, 
and can be done analytically.  We obtain
\begin{eqnarray}
\tilde{m}^2 &=& 
2 \zeta(3) \Gamma(4) C(i) \frac{g_{(5)}^2}{16 \pi^2} \frac{F^2_\Sigma}{M_*^5 L^4}
\label{scalars-eq}
\end{eqnarray}
There are two things to notice about this result.  First, there is a 
numerical enhancement of the coefficient partially due to the sum of
Kaluza-Klein (KK) modes.  Second, the form of the contribution
Eq.~(\ref{scalars-eq}) looks similar to that of gauge mediated supersymmetry 
breaking for scalars \cite{GMSB}.
The main difference is that the above contribution comes at 
one-loop with $g^2/16 \pi^2$ and not two loops with $(g^2/16 \pi^2)^2$.  
The gauge couplings, however, are unified at this scale and thus 
the ratio of squared scalar masses ${\tilde m}_i^2$ and ${\tilde m}_j^2$ 
is simply $\sum_a C_a(i)/\sum_b C_b(j)$ as it would be for 
gauge mediation at the GUT scale.

Using $m_{3/2} = F_{\Sigma}/M_{\mathrm{Pl}}$, we can rewrite the expression
for the scalar masses in Eq.~(\ref{scalars-eq}) in terms of the
gravitino mass
\begin{eqnarray}
\tilde{m}^2 &=& 
2 \zeta(3) \Gamma(4) C(i) \frac{g^2}{16 \pi^2} \frac{1}{(M_* L)^2} m_{3/2}^2 .
\label{gaam-eq}
\end{eqnarray}
As predicted in our earlier estimate, this ``gaugino-assisted'' 
contribution to the scalar masses 
is suppressed by one loop factor and two volume factors
$M_* L$ with respect to $(m_{3/2})^2$.  In contrast, 
the anomaly-mediated contribution is simply two-loop suppressed.  
If the operator in Eq.~(\ref{gauginokinetic}) has a coefficient of order one, 
the two contributions are of the same order if 
$(M_* L)^{-2} \sim g^2/(16\pi^2)$.
In this case, the slepton mass problem of AMSB is solved.
We need only require that the sign of the operator in 
Eq.~(\ref{gauginokinetic}) is such that the 
contribution to scalar (mass)$^2$ is positive.  Since we
are simply writing the effective operators, the coefficient,
including the sign, is undetermined.


All of the operators in Eqs.~(\ref{opsb})--(\ref{opsf}) 
give contributions to scalar masses of the same form as 
in Eq.~(\ref{gaam-eq}).  The operator Eq.~(\ref{opsc}) contributes
to both $(F_{\mu\nu})^2$ and $D^2$ terms\footnote{We ignore the
$F\widetilde{F}$ term as we assume CP is conserved on the hidden sector 
boundary.  This allows us to solve the supersymmetric CP problem 
\cite{FCNC,CLNP}, but not the strong CP problem.}.  The $(F_{\mu\nu})^2$
term contributes through the diagrams in Fig.~\ref{loop}(b) and \ref{loop}(c).
In the five-dimensional theory, the $D^2$ term is converted to 
$(X^3 - \partial_y \Phi)^2$, where $X^3$
and $\Phi$ are real auxiliary and scalar components of the five-dimensional
vector multiplet, respectively \cite{MP}.  This replacement reveals
the fourth diagram, Fig.~\ref{loop}(d).  Using the scalar propagator in 
Ref. \cite{MP}, we get
\begin{equation}
\delta m_{\tilde{f}}^2 = 2 \frac{g_{(5)}^2 F_{\Sigma}^2}{M_*^5} 
   C(i) \int \frac{d^4 q}{(2\pi^4)}\left(\frac{1}{L}\sum_{n=-\infty}^{\infty}
   \frac{q_5^2}{q^2 + q_5^2}\cos{q_5 L}\right)^2 \frac{1}{q^2},
\end{equation}
where $q_5\equiv n\pi/L$ and all diagrams are calculated in Euclidean 
space.  The sums in this integral are asymptotic series.  They become 
convergent if we make the assertion
\begin{equation}
0=\sum_n (-1)^n = \sum_n (-1)^n \frac{(q^5)^2 + q^2}{(q^5)^2 + q^2},
\end{equation}
which allows us to replace the $(k^5)^2$ terms in the numerator with $k^2$.
This was done in \cite{MP} and shown to be a necessary ingredient to preserve
supersymmetry.  Making this replacement, we get a result of identical form
as Eq.~(\ref{scalars-eq}).  

The remaining soft parameters to compute are scalar trilinear, or
`A' terms.  Scalar trilinear couplings do not get contributions from the 
operators (\ref{opsb})-(\ref{opsf}) and are non-zero at the compactification
scale solely due to anomaly mediation \cite{RS,GLMR}.
This is clear since these operators preserve an $R$ symmetry which A terms
break.  Other operators produce at most negligible contributions.

A few comments about the spectrum are in order.  In the four-dimensional
description, there is a tower of KK mode copies of the gauge multiplets 
which all feel supersymmetry breaking from the conformal compensator.
Normally we should expect these states to completely decouple as is 
typical in AMSB.  However, it is non-trivial to show that this generically
happens as KK masses come directly from dynamics of fields which 
stabilize the extra dimension.  We shall assume that these effects can be made 
negligible in some cases and leave explicit calculations for future work.

The other threshold one might worry about is that of the GUT scale.
If there is a unified theory, there are additional matter fields
with GUT scale masses such as Higgs triplets (or the remainder of the
Higgs multiplet) or the fields that break the GUT group.  These are chiral
superfields with a supersymmetric mass of $\sim \mg$.  From
the superconformal anomaly, they will get an order weak-scale holomorphic 
soft mass squared 
of order $m_{3/2}\mg$.  They will also get a non-holomorphic soft mass
squared from the one-loop contributions described above.  When these 
fields are integrated out, their holomorphic
supersymmetry breaking effects on matter scalar soft masses decouple 
\cite{RS,GLMR,PR,CLMP,KSS}.  The non-holomorphic piece does not 
contribute to gaugino masses at leading order \cite{PT,kribsgaugino} 
and as it is of order the weak scale, its contribution to scalars 
is \emph{at least} one-loop suppressed compared to the leading 
contributions\footnote{Note, however, that if the GUT scale is generated 
dynamically through a mechanism analogous to that of Pomarol and Rattazzi, 
then the GUT physics does \emph{not} decouple, in some cases leading
to a viable spectrum \cite{CLPSS}.}.

One impact the GUT fields could have is on the boundary conditions of
the MSSM scalars if $L^{-1}\ll M_{\rm GUT}$.  We discuss this possibility 
in the next section and in the Appendix.

To summarize, we have found a model that has a gaugino spectrum 
generated purely from anomaly mediation, while the scalar spectrum 
results from summing two contributions to their squared masses:  
the pure anomaly-mediated contribution plus a gaugino-assisted
contribution that looks like gauge mediation at the GUT scale.  
The latter can be positive and thus the slepton mass problem is solved.  

\section{Spectrum and phenomenology}
\label{pheno}

We have seen that gaugino mediation without singlets yields 
a viable model with large contributions resulting from
anomaly mediation.  Since the gaugino masses are generated
exclusively from anomaly mediation while the scalar masses
have both the anomaly contributions as well as the 
gaugino-assisted contributions that we found above, 
the spectrum of the model has qualitative similarities with
the phenomenological fix of adding a universal scalar mass squared 
to the anomaly-mediated contribution at the unification scale \cite{GGW},
which we call ``$\mathrm{AMSB} + m_0^2$'' scenario.  However, it is 
clear from Eq.~(\ref{gaam-eq}) that our ``additional'' scalar mass 
contribution is \emph{not} universal, but instead proportional to 
a weighted sum over the quadratic Casimirs for the matter field of interest
\begin{eqnarray}
\tilde{m}_i^2 \propto \sum_a C_a(i) g_a^2 \qquad \qquad a \; = \; 
U(1)_Y, SU(2)_L, SU(3)_c \; .
\end{eqnarray}
In the case where the size of the extra dimension is of order
the unification scale, the gauge couplings are the same 
and can be factorized out of the above.  The gaugino-assisted scalar 
mass contributions therefore differ merely by a sum over the quadratic 
Casimirs for each representation, which we show 
in Table~\ref{Casimir-table}.
\TABLE[t]{
\begin{tabular}{c|cc} \hline\hline
Representation & SM Casimir weight & $X$ and $Y$ gauge boson Casimir weight 
\\ \hline
  $Q$ & $21/10$ & $3/2$ \\
  $u$ & $8/5$   & $2$ \\
  $d$ & $7/5$   & $1$ \\
  $L$ & $9/10$  & $3/2$ \\
  $e$ & $3/5$   & $3$ \\
$H_u$ & $9/10$  & $3/2$ \\
$H_d$ & $9/10$  & $3/2$ \\ \hline\hline
\end{tabular}
\caption{The Casimir weights that enter the gaugino-assisted
contribution to scalar masses resulting from five-dimensional loops with 
ordinary SM gauge fields (middle column) and $X$ and $Y$ gauge boson 
fields of SU(5) (right column), for each SM representation.  
Note that we have assumed $H_u$ and
$H_d$ are in the fundamental representation of SU(5).}
\label{Casimir-table}
}


Gaugino-assisted anomaly mediation has only a small number of input 
parameters in addition to those of the standard model, and is thus 
a highly predictive model of new physics.  The anomaly-mediated 
contribution to soft terms is essentially dictated by a single unknown
parameter, $m_{3/2}$.  The gaugino-assisted contribution comes
from the operators in Eq.~(\ref{opsb})--(\ref{opsf}) and are parameterized 
by unknown effective theory coefficients.  Since the contributions 
from all of these operators are identical up to each of their order 
one coefficients, we parameterize their contributions by a single 
order one coupling $\eta$ which multiplies the quantity in 
Eq.~(\ref{gaam-eq}).
In addition, in the absence of unification, there are really three 
sets of operators with in principle differing coefficients, 
corresponding to the three gauge groups of the SM.  We shall assume 
that gauge coupling unification is not an accident, and therefore the 
coupling is the same for each gauge group.  This is a necessary condition 
for embedding our model within a GUT group.

A third input of this model is the volume factor $M_* L$.  We restrict 
ourselves to compactification scales which do not disrupt normal 
four-dimensional gauge coupling unification, \emph{i.e.}, we require
$L^{-1} \gsim \mg$.  This also means that we have no new proton decay
problems beyond those of ordinary (four-dimensional) supersymmetric models.
Using Eq.~(\ref{add}), this means there is an upper bound $M_*L\lsim 22$.
We are anyway interested in keeping $M_*L$ small so as not to introduce
a new large hierarchy.  However, as we explained in Sec.~\ref{amsb},
avoiding bounds on FCNC requires $M_*L\gsim 16$, or equivalently, 
$L^{-1} \lsim 2 \mg$.  

Changing $M_*L$ affects the size of gaugino-assisted contribution,
Eq.~(\ref{gaam-eq}), but this model dependence can be equivalently absorbed 
in the coefficient $\eta$, and thus we can simply set $L^{-1}$ to $M_{\rm GUT}$
(if we ignore the small effects of the additional running).  However, there 
can be an additional effect from decreasing $M_*L$, which is the turning 
on of contributions from GUT fields.  Hence, we must distinguish between 
gaugino-assisted anomaly mediation with or without GUT field contributions.  
In the case where we do have a GUT at the unification scale, we find that
the GUT contributions are significant for the entire range of $L^{-1}$.
In fact, for $L^{-1}$ near its upper limit of $2 M_{\rm GUT}$, 
gaugino-assisted contributions become nearly universal over complete 
GUT multiplets.  Thus, for grand-unified gaugino-assisted anomaly mediation, 
an additional parameter $r=L^{-1}/M_{\mathrm{GUT}}$, with $1\lsim r\lsim2$,
must be added.  There is, of course, the additional issues of
which GUT group, the representation that contains the Higgs, etc.

Finally, the remaining parameters are those of the Higgs sector.  The
magnitude of the supersymmetric mass parameter $\mu$ is fixed by the
measured $Z$ boson mass while the soft parameter $B$ is unknown.
We follow the usual practice and exploit the other electroweak symmetry 
breaking condition to trade the high scale parameter $B$ for the weak scale 
parameter $\tan\beta = \langle H_u \rangle/\langle H_d \rangle$.  

Thus the model has the following free parameters:
\begin{equation}
m_{3/2},\: \eta,\: \tan\beta,\: {\rm and \: sign}(\mu),
\end{equation}
and also (at least) one more parameter, $r$, that enters if a GUT 
exists at the unification scale.  We should emphasize that the above does 
not incorporate a specific mechanism to generate the $\mu$ term.  As a result,
we should also add the parameters $\delta m_{H_u}^2$ and 
$\delta m_{H_d}^2$ which are contributions to the Higgs soft parameters 
over and above those coming from AMSB and the gaugino-assisted
contributions shown in Fig.~\ref{loop}.  In principle, these additional 
Higgs mass contributions could result not just from a $\mu$-term
mechanism (we discuss possible mechanisms to generate the
$\mu$ term in the next section) but also from GUT threshold effects, 
\emph{e.g.}, our ignorance of the representation 
in which the Higgs field is embedded and/or the additional multiplets 
required for some version of doublet-triplet splitting.  
For simplicity, in the results we present below we shall assume
that $\delta m_{H_u}^2$ and $\delta m_{H_d}^2$ are negligible, however
a more thorough analysis of the parameter space is warranted.

\FIGURE[t]{
\epsfxsize=4.5in
\centerline{\epsfbox{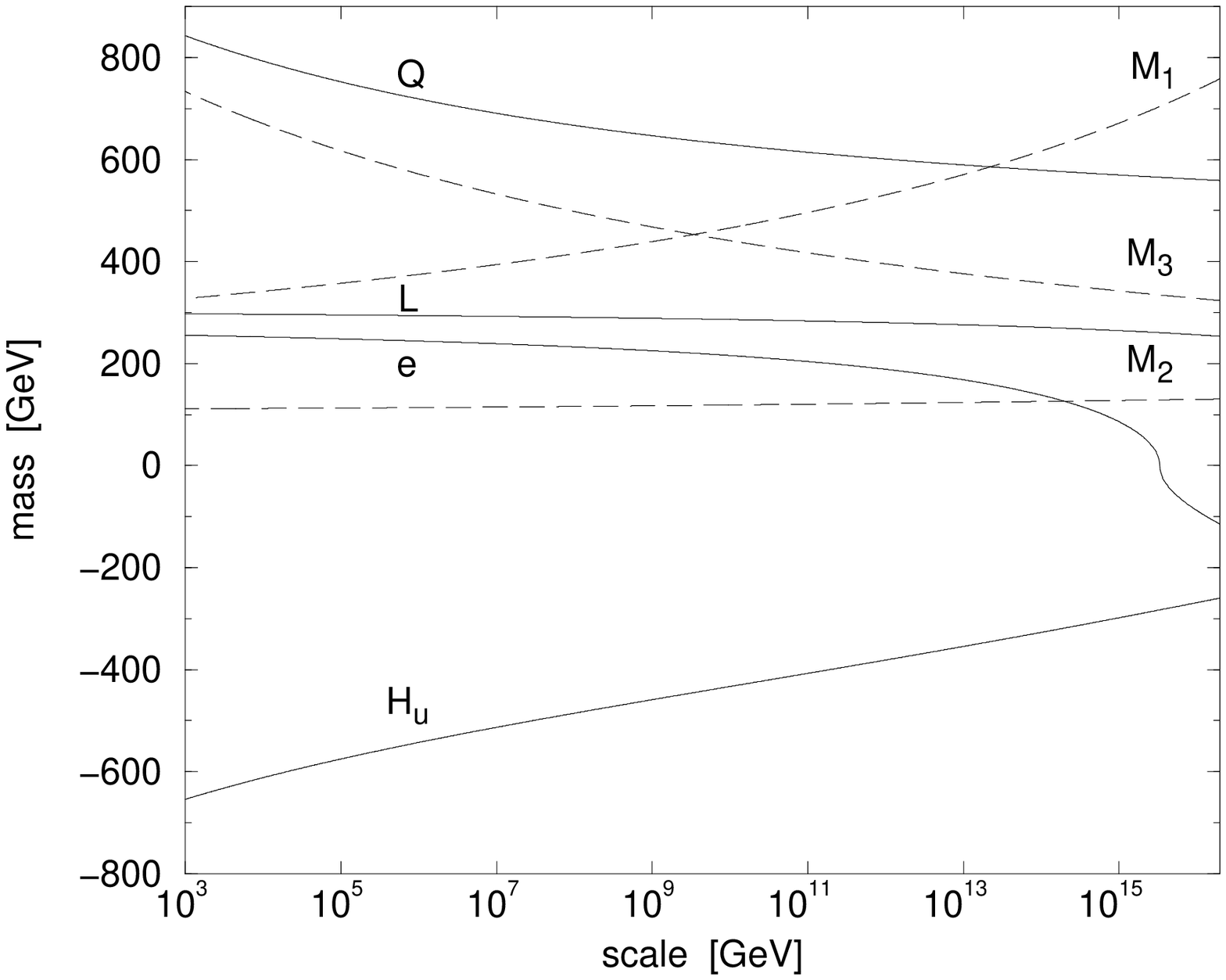}}
\caption{Soft masses as a function of scale.  The input parameters are
as follows:  $m_{3/2}=35$ TeV, $\eta=1$, $\tan\beta=5$, $\mu > 0$,
and $r = 1$.  We assume no new fields exist at the scale $M_{\rm GUT}$.
For the gauginos (dashed) we plot $|M_i|$, and for scalars (solid) we plot 
sign$(m^2) \sqrt{|m^2|}$.}
\label{RGE}}

Using the above set of input parameters, we can now compute the low energy 
spectrum.  At the unification scale, the AMSB contribution to the scalar
masses is summed with the gaugino-assisted contribution, and the
entire set of renormalization group (RG) equations is evolved to the
weak scale.  We use full two-loop RG evolution in gauge couplings,
Yukawa couplings, gaugino masses, scalar masses, and scalar trilinear
couplings.  
In Fig.~\ref{RGE}, we illustrate the running of soft masses 
with the input parameters $m_{3/2}=35$ TeV, $\eta=1$, $\tan\beta=5$ and
$\mu > 0$, and without a GUT.  Notice that the squarks are the 
heaviest sparticles of the spectrum, followed by the gluino, some combination 
of the Bino, sleptons, and Higgs, and finally the Wino as the LSP\@.  These 
features are generic to most of the allowed parameter space.  Since we are 
adding positive contributions to all of the sparticles, it is important to 
note that the up-type Higgs (mass)$^2$ is negative, and therefore electroweak 
symmetry is broken radiatively as usual.  The gaugino masses are precisely 
proportional to the SM beta functions, identical to the spectrum studied in the 
$\mathrm{AMSB} + m_0^2$ scenario.

\FIGURE[t]{
\centerline{\hfill \epsfxsize=0.60\textwidth
\epsffile{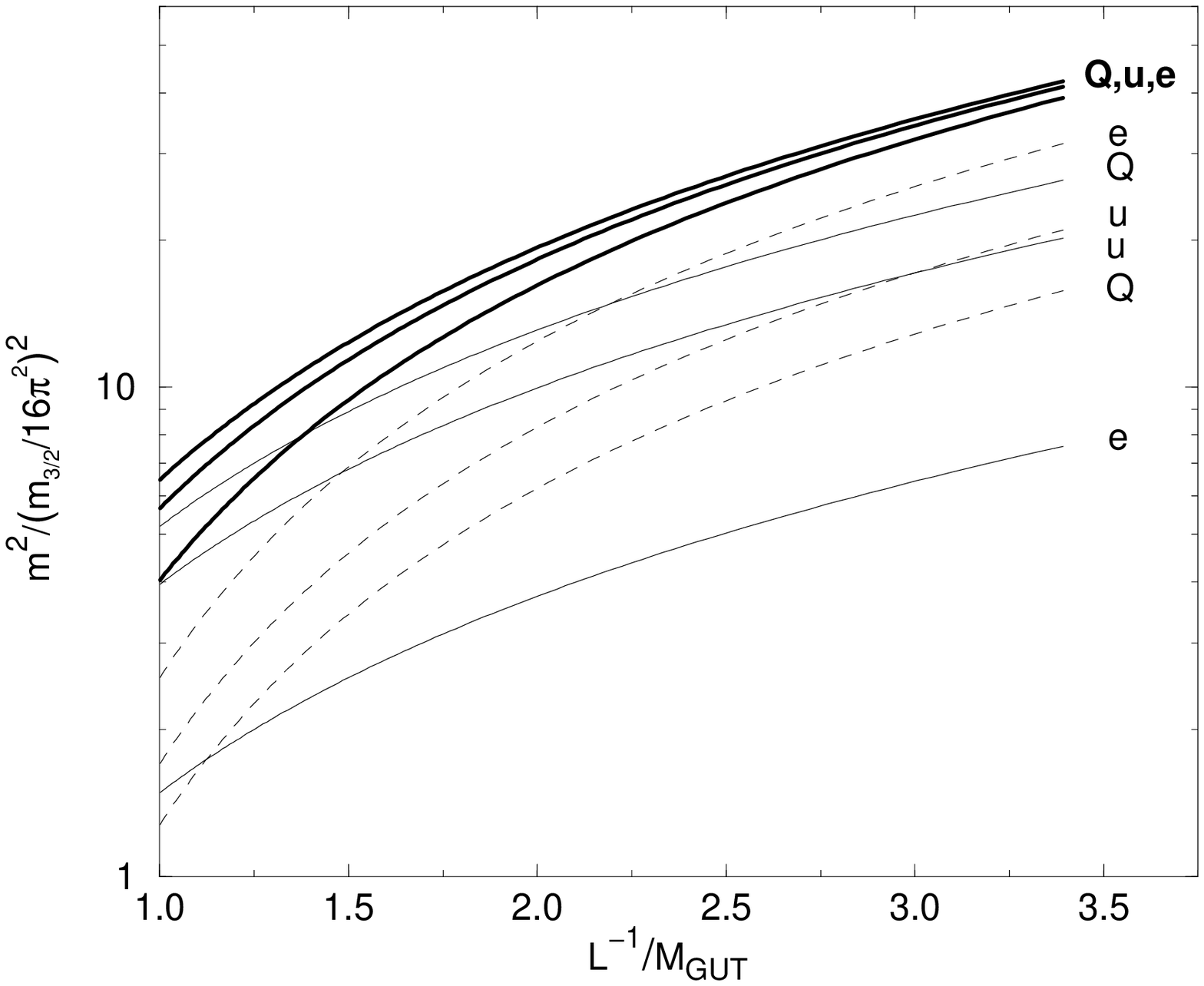}
\hfill \epsfxsize=0.60\textwidth
\epsffile{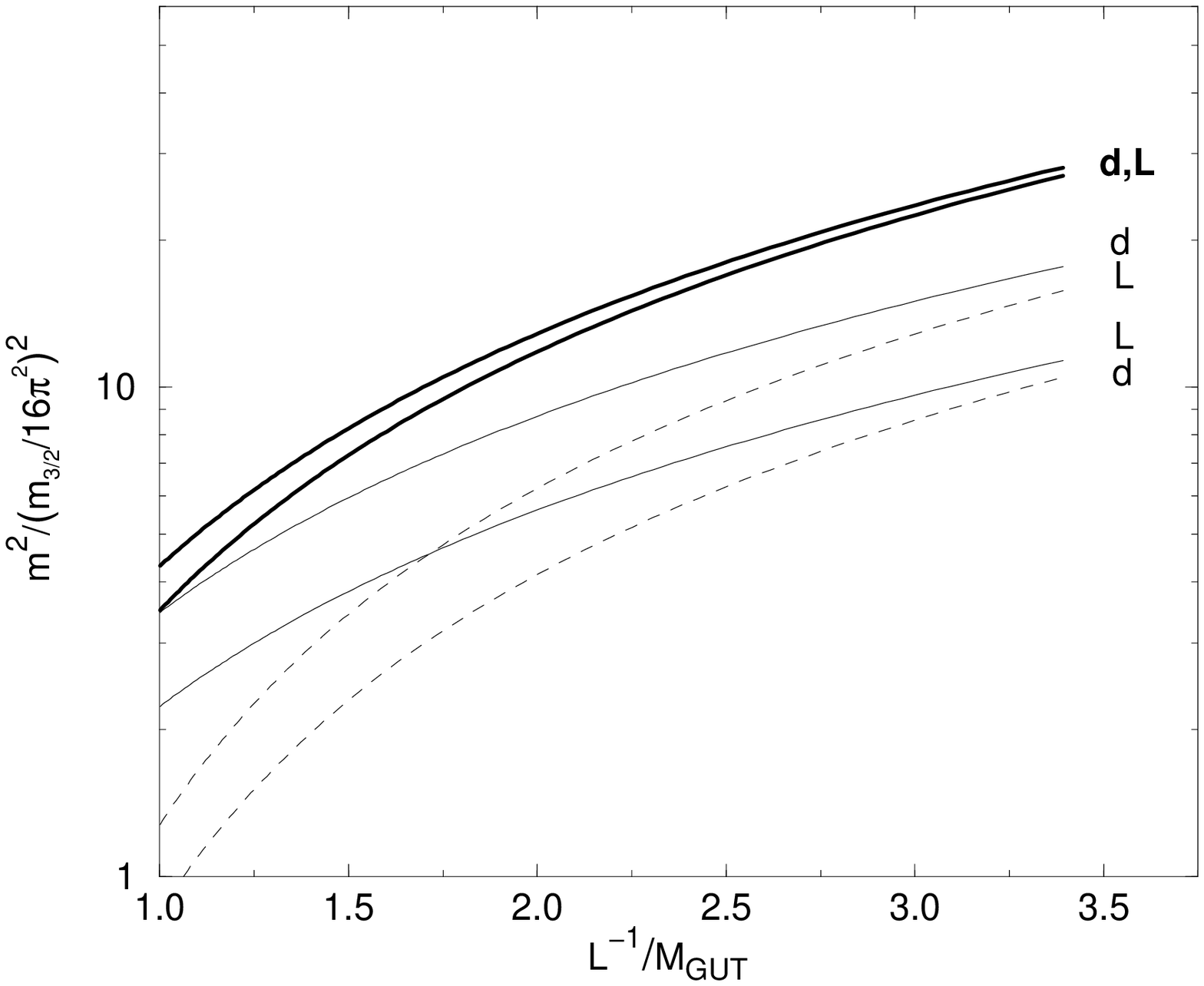}
\hfill }
\caption{Gaugino-assisted contributions to squared scalar masses in units of 
$m_{3/2}/16\pi^2$.  The thin solid lines indicate the contribution from
SU(3)$\times$SU(2)$\times$U(1) gauge multiplets, the dashed lines 
represent contributions from $X$ and $Y$ gauge bosons and gauginos of 
the complete SU(5), and the thick solid lines are the sum of
both contributions.  For $r\gg 1$, the sums of these contributions
converge for each complete multiplet ($Q$, $u$ and $e$ of the $10$ and 
$d$ and $L$ of the $\overline{5}$.)}
\label{GUT-fig}}

Now let's look at a spectrum assuming the existence of an SU(5) GUT.  
It is important to take into account the full gauge multiplet of the unified
theory, as the additional gauge fields will now appear in the loops in
Fig.~\ref{loop}.  We compute this contribution in the Appendix for $r\geq1$
in the case of an SU(5) GUT and present the results in Fig.~\ref{GUT-fig}.  
Notice that the contributions are quite significant for both the left-handed 
and right-handed sleptons.  In fact, even at $r=1$ the contributions to 
the right-handed sleptons exceeds that from the standard model gauge multiplets.  

\FIGURE[t]{
\centerline{\hfill \epsfxsize=0.60\textwidth
\epsffile{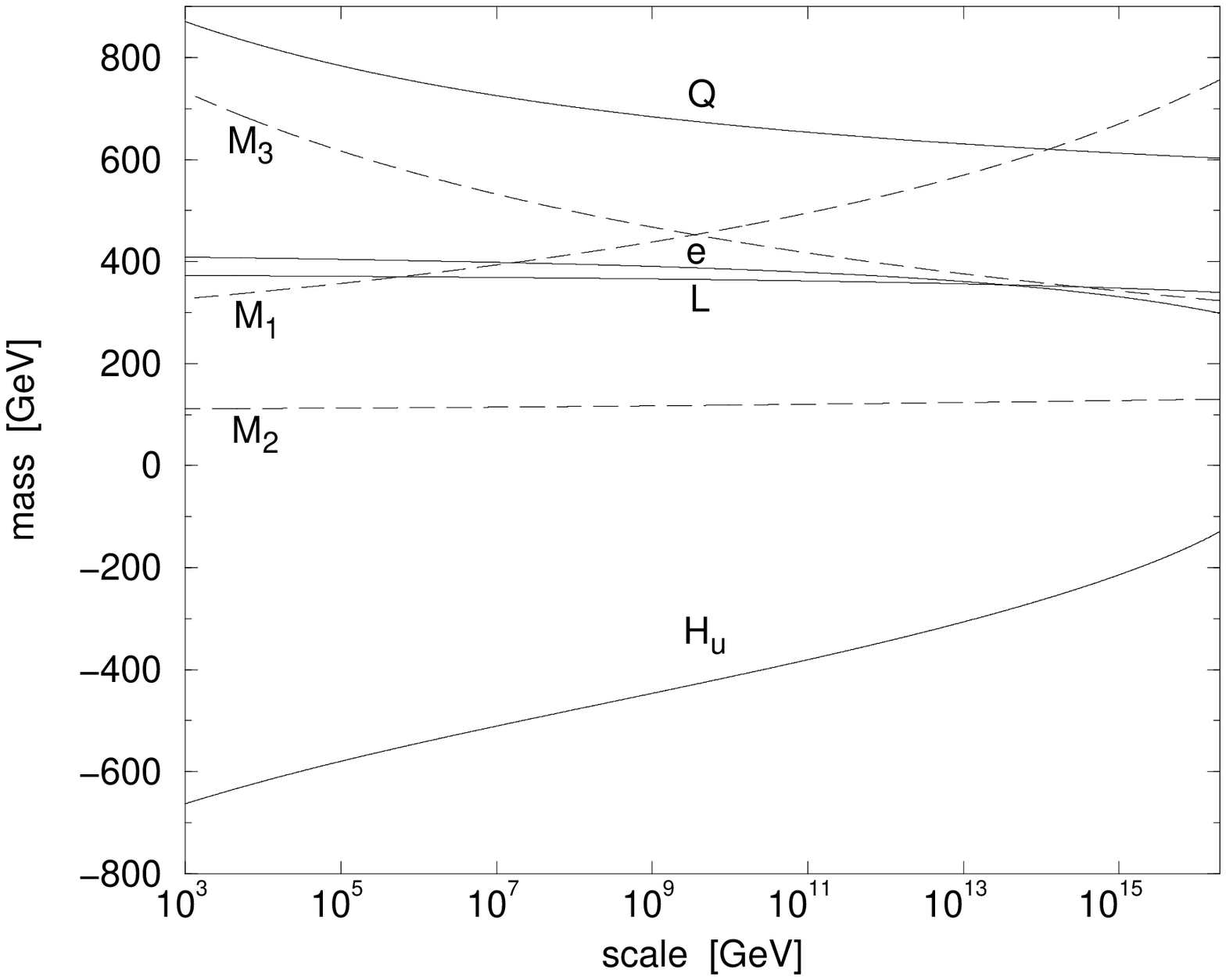}
\hfill \epsfxsize=0.60\textwidth
\epsffile{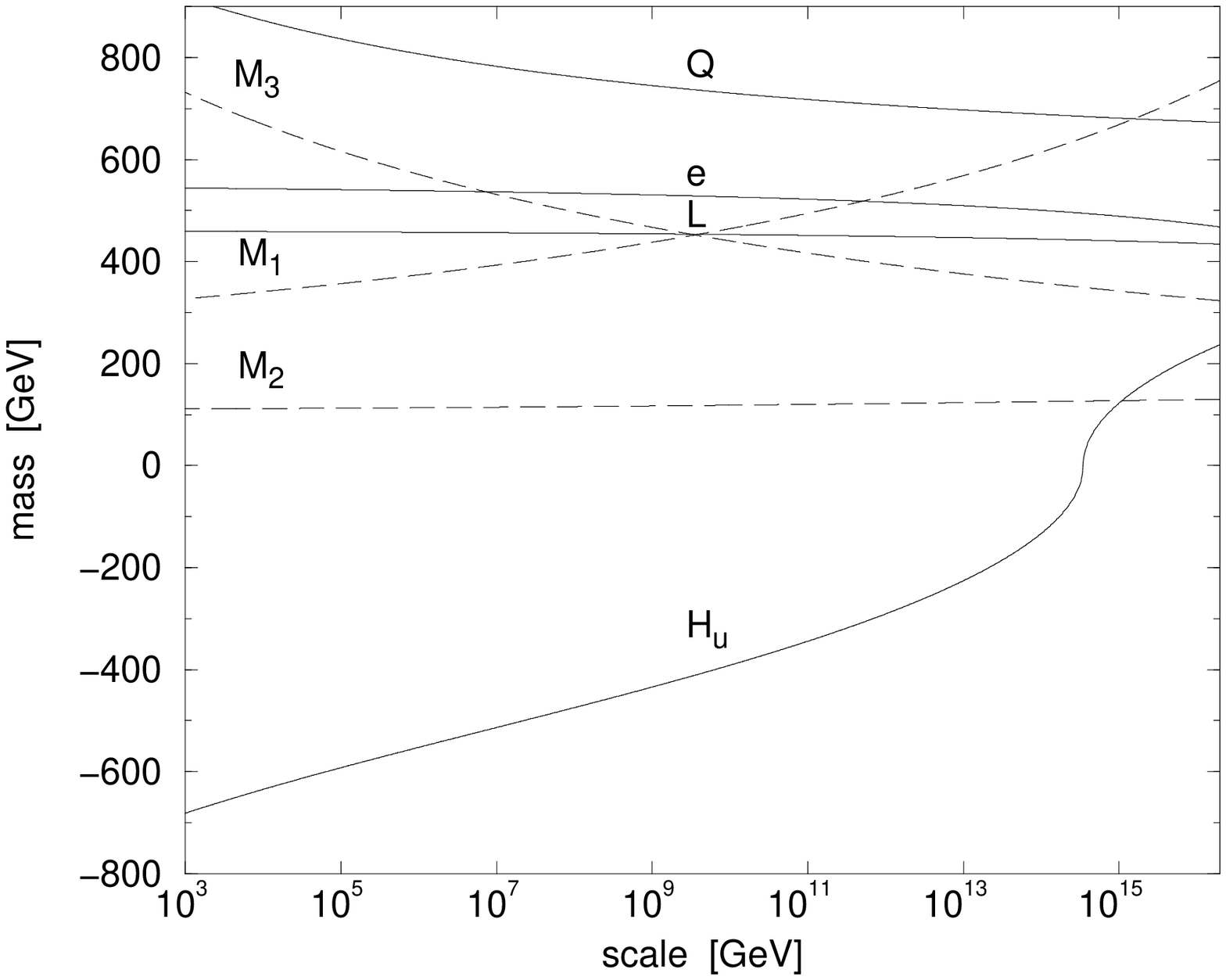}
\hfill }
\caption{Soft masses as a function of scale.  The input parameters 
for the figure on the left are the same as in Fig. \ref{RGE} (with
 $r = 1$).  The figure on the right is the same except with 
$\eta=0.5$ and $r = 2$.  A minimal SU(5) GUT is assumed to exist at 
$M_{\rm GUT}$.  As before, the gauginos (scalars) are the dashed 
(solid) lines.}
\label{RGE_GUT}}

To see the effects of the GUT on the spectrum, we run two sets of 
parameters, one with $m_{3/2}=35$ TeV, $\eta=1$, $\tan\beta=5$, 
$\mu > 0$ and $r=1$, and another with the same inputs except $\eta=1/2$
and $r=2$.  The results are presented in Fig.~\ref{RGE_GUT}.  Note the most 
significant change is that the left-handed sleptons are now the lightest
scalar superpartners in all of our parameter space.

\FIGURE[t]{
\epsfxsize=3.5in
\centerline{\epsfbox{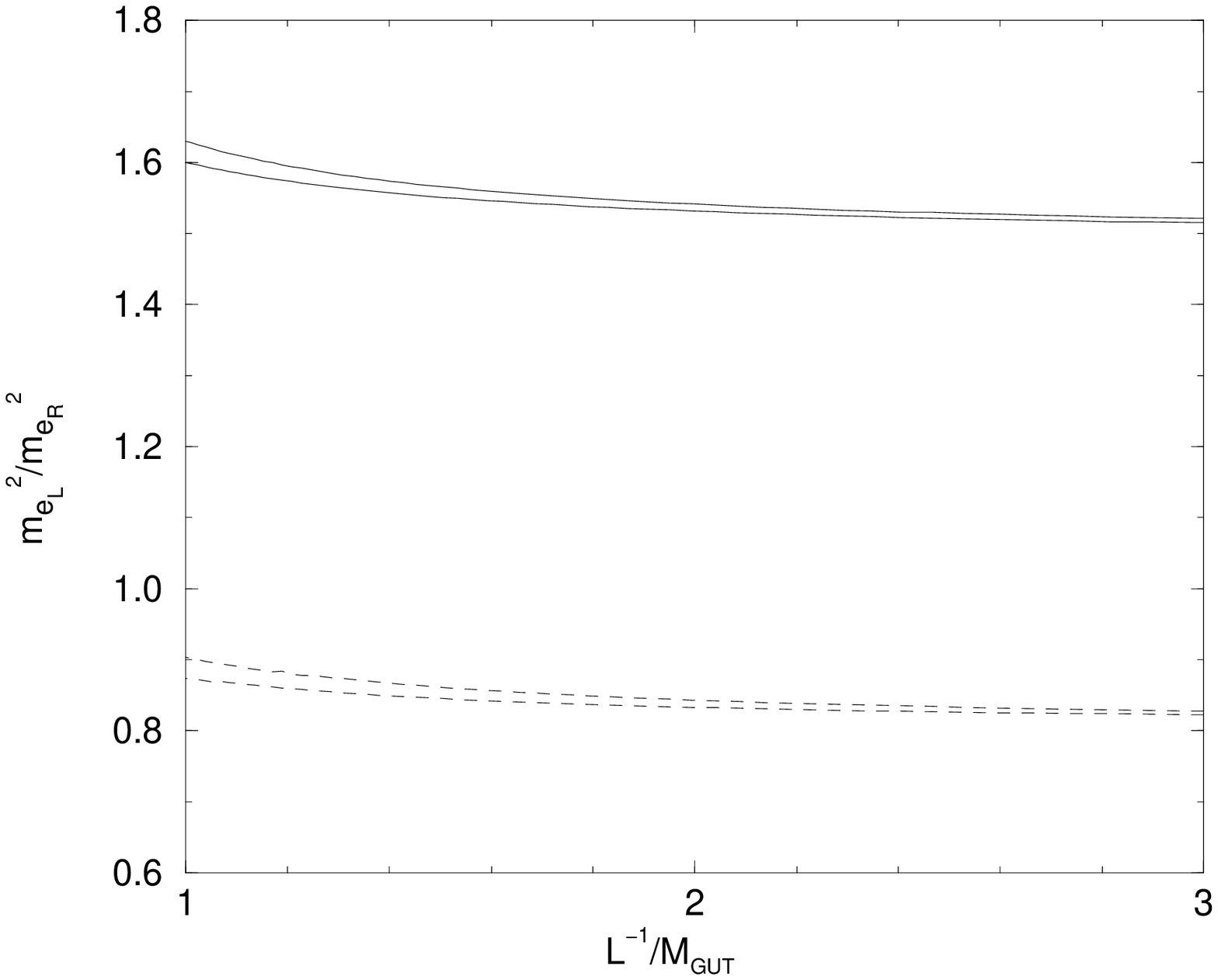}}
\caption{Ratio of the left-handed to right-handed slepton masses
as a function of $L^{-1}/M_{\mathrm{GUT}}$ evaluated at the weak scale.  
The (upper, lower) solid lines correspond to models \emph{without} GUT 
contributions, while the (lower, upper) dashed lines correspond to models 
with SU(5) GUT contributions (including thresholds), for 
$m_{3/2} = (35, 70)$~TeV\@.  Note that in the AMSB $+ m_0^2$ scenario, the 
ratio is $1$ to within a few percent \cite{GGW}, while in generalized 
gauge-mediation the ratio is significantly less than one \cite{KribsAM}.}
\label{slepton-fig}}

Another important consequence of the non-degeneracy of the 
gaugino-assisted scalar mass contribution is that the left-handed
and right-handed slepton masses are no longer accidentally degenerate.
It was first shown in Ref.~\cite{GGW} that the spectrum of AMSB 
with the phenomenological fix of adding a universal $m_0^2$ implies
the weak scale masses $m_{\tilde{L}}$ and $m_{\tilde{e}}$
are the same to within a few percent.  This is one of the few predictions 
that can also be used to distinguish this framework from generalized gauge 
mediation \cite{KribsAM}.  In gaugino-assisted anomaly mediation,
however, the gaugino-assisted contributions to the slepton masses 
differ by the ratio of their summed quadratic Casimirs.
This can be read off from Table~\ref{Casimir-table},
$\delta m_{\tilde{L}}^2/\delta m_{\tilde{e}}^2 = 3/2$.
Since the anomaly-mediated contribution is also larger (but negative)
for the left-handed slepton mass, one must combine the above with the 
anomaly-mediated contributions and evolve to the weak scale to 
determine the slepton mass spectra.  In Fig.~\ref{slepton-fig}
we show the ratio of the left-handed to right-handed slepton masses
at the weak scale as a function of $L^{-1}/M_{\mathrm{unif}}$, 
both with and without heavy GUT field contributions.
Note that, with or without a GUT, this ratio is nearly 
independent of the other parameters of the model, including the 
size of supersymmetry breaking $m_{3/2}$.  
When GUT field contributions are not present, we see 
that the left-handed slepton mass is always significantly
larger than the right-handed slepton mass.  Once GUT field
contributions are included, the ratio is less than one. 
However, this last statement depends on the GUT group chosen. 
In fact, if we instead choose SO(10), the contributions to all
chiral matter will be essentially universal (at large enough $r$) and
the sleptons would be nearly degenerate, as occurs in the spectrum
of $\mathrm{AMSB} + m_0^2$ (though for very large $r$ one should take
into account the renormalization group running above the GUT scale).

Since the gaugino masses are proportional to the SM beta functions,
several analyses of the phenomenology and signals performed in 
the $\mathrm{AMSB} + m_0^2$ scenario apply here as well \cite{GGW, FM}.  
In particular, the lightest neutralino is mostly a Wino, with the 
lightest chargino nearly degenerate in mass.  A careful calculation of
this mass splitting was first done in \cite{CDM}.  For a Wino LSP, the 
near degeneracy results in a macroscopic decay length for the lightest 
chargino, typically of order a few centimeters \cite{CDM, FMRSS, GGW} 
implying unique experimental signatures which have been analyzed by a 
number of groups \cite{FMRSS,GGW,FM,KribsAM,AMSBpheno}.  In addition, the model 
also shares the cosmological features of the $\mathrm{AMSB} + m_0^2$ scenario, 
including relaxing the cosmological problem associated with gravitino
decay during nucleosynthesis, and the possibility of Wino LSP dark 
matter produced via non-thermal primordial gravitino decay \cite{GGW, MR}.

\section{Additional bulk fields and the $\mu$ term}
\label{bulk}

No model of supersymmetry breaking would be complete without a mechanism
for generating the supersymmetric mass parameter $\mu$ for the superpotential
operator $\mu H_u H_d$.  As $\mu$ should be of the same order as soft masses,
it is natural to assume that the $\mu$ term is generated by supersymmetry
breaking.

There are several mechanisms on the market \cite{RS,PR,KSS,CLMP} specific
to AMSB.  However, maintaining the attractive features of the model restricts
which mechanisms we can use.  We discuss the restrictions, putting emphasis on the 
effects of adding bulk fields, and point out the class of mechanisms that
work in our context.

The first requirement is that the Higgs fields live on the matter sector
boundary.  If the Higgs fields live in the bulk, their squared soft masses
are expected to be of order $m^2\sim m_{3/2}^2\gg m_{\tilde q}^2$ due to
direct couplings at the hidden sector boundary, and thus we avoid this 
scenario.

In addition, light bulk chiral superfields, Higgs or otherwise, can give 
significant enough contributions to flavor violation to encroach on current
FCNC bounds.  To see this, we introduce a bulk chiral superfield $B$.
Now, the following operators can appear in the Lagrangian:
\begin{equation}
\int d^4\theta \frac{\Sigma^{\dagger}\Sigma}{M_*^3} B^{\dagger}B \delta(y-L)
	\:\:\:\:{\rm and}\:\:\:\: \int d^4\theta 
	\frac{Q_i^{\dagger}Q_j}{M_*^3} B^{\dagger}B \delta(y) \label{opbulk}
\end{equation}
We can easily estimate $B$'s one-loop contribution to squared scalar masses.
It is proportional to $F_{\Sigma}^2/M_*^6$ and the effective cutoff of
the loop integral is $L^{-1}$, so dimensional analysis gives us an estimate
of the scalar mass contribution to be
\begin{equation}
\delta m_{ij}^2 \sim 
	\frac{1}{16\pi^2}\frac{F_{\Sigma}^2}{M_*^2}\frac{1}{(M_* L)^4}
	\sim \frac{1}{16\pi^2}m_{3/2}^2 \frac{1}{(M_* L)^3}
\end{equation}
which is only a volume factor suppressed when compared with the dominant
scalar mass contribution.  With coefficients of order one and numerical 
factors taken into account, this contribution is about an order of magnitude
too large for $i,j=1,2$ \cite{FCNC}.  One can either forbid new bulk
chiral superfields or simply require that their couplings to boundary fields
are $\lsim {1\over 3}$\footnote{We note that in gaugino mediation \cite{KKS,CLNP}, 
these contributions are below experimental bounds as the flavor blind scalar 
masses are of order $\sim m_{3/2}$.},  allowing $\mu$-term mechanisms such as 
``shining'' \cite{AHHSW,SS1}.  Vector multiplets in the bulk allow operators 
similar to those above but suppressed by additional powers of $M_*$.  The additional 
volume suppression renders them harmless with respect to FCNC.

Viable mechanisms for producing the $\mu$ term that do not require
additional bulk fields or contact interactions between the matter and
hidden sectors appear in \cite{PR,KSS}.  These models require
the existence of one or more standard model singlets to live on the matter
boundary.  For example, a $\mu$ (and $B \mu$) term of order the weak scale can 
be generated from an operator $H_u H_d X^{\dagger}/X$ in the 
K\"{a}hler potential.  This operator gets a coefficient 
$\tilde{Z}\left(\sqrt{XX^{\dagger}/\Phi\Phi^{\dagger}}\right)$ from
wave-function renormalization, where $\sqrt{XX^{\dagger}/\Phi\Phi^{\dagger}}$
should be taken as the renormalization scale and $X$ is the modulus-like 
field which has a large scalar vev but negligible auxiliary component.
As shown in \cite{PR}, this operator generates a $\mu$ term at one loop and
a $B\mu$ term at two loops.  The operator can be generated by a superpotential
$\left[\lambda S H_u H_d + k S^3 + y S^2 X\right]$ (where $S$ is a singlet)
and a kinetic mixing term between $X$ and $S$.

Finally, we comment on stabilization of the compactified dimension.  
As in anomaly mediation, the compact dimension can be stabilized by the
mechanisms of \cite{LS} and \cite{AHHSW}.
However, a very recent analysis by Chacko and Luty \cite{RMSB} 
suggests that additional contributions to gaugino masses arise in the
scenario of Ref.~\cite{LS} from the radion multiplet if the gauge multiplets are 
in the bulk.  The stabilization mechanism of \cite{AHHSW} works in our models 
without significantly affecting the spectrum as long as one assumes slightly 
suppressed couplings of the new bulk fields to the boundaries as outlined above.  
It is an interesting prospect to see if this mechanism (or any new one) can be 
embedded into supergravity.  We leave this speculation for possible future work.

\section{Conclusions}
\label{conclusion}

We have presented a model of mediating supersymmetry breaking through
an extra dimension.  By putting standard model gauge fields and their
superpartners in the bulk while requiring the hidden sector to be free of 
singlets, the gaugino spectrum is exactly that of anomaly-mediated 
supersymmetry breaking \cite{RS,GLMR}.  The scalar masses obtain contributions 
from both anomaly mediation and operators which appear generically on the 
hidden-sector boundary.  The latter amounts to non-supersymmetric
contributions to wave-function renormalization of the gauge multiplet
inducing scalar masses at the one-loop level.  For order one couplings, 
the gaugino-assisted contribution is more than sufficient to make 
the squared masses of the sleptons positive.

The operators introduced in Sec.~\ref{gaam-sec} induce
a hard breaking of supersymmetry in the four-dimensional effective theory
below the scale $L^{-1}$.  
The hard breaking manifests itself as a tiny difference between
the gauge and gaugino couplings to matter resulting in
quadratically divergent contributions to scalars at one loop.
We find that this way of calculating the contribution gives the
same functional result as an explicit five-dimensional loop calculation.
Furthermore, these operators should appear in any general 
four-dimensional effective theory with a hidden sector since no
symmetries (including $R$ symmetries) are broken by them.  The 
contributions to scalar masses from these operators are one-loop 
suppressed compared to those that come from contact terms.  Thus, they 
are important when contact terms are absent, such as in models with 
sectors separated spatially in extra dimensions.

The phenomenology of the model contains some of the interesting features
of the anomaly-mediated spectrum with a universal mass squared $m_0^2$ added
to scalar fields \cite{GGW}.  In fact, if all standard model matter lives
in a single multiplet of a grand unified theory, the contribution from
gaugino-assisted anomaly mediation could be {\it precisely} a universal
scalar mass.  However, if you assume no new physics at the GUT scale, 
or a unified group like SU(5), SU(6) or SO(10), the new contributions
would not be universal, thus offering new spectra and thus new phenomenology.
We therefore believe a more thorough analysis of this model would be of
interest, as it is a new, highly predictive and very simple way of mediating 
supersymmetry breaking.  

\appendix

\section*{Appendix: A GUT threshold calculation}

In this Appendix, we make explicit the calculation of gaugino-assisted contributions
to scalar masses for compactification scales at or above the GUT scale.  We assume
there exists a unified theory and thus the loop contributions from additional gauge
fields can be important.  For concreteness, we assume a minimal SU(5) GUT and then 
discuss the model dependent and independent features of our results.

As discussed in the text, the compactification scale $L^{-1}$ cannot be too much
larger than $M_{\rm GUT}$.  Therefore, in calculating the contributions to scalar
masses from $X$ and $Y$ bosons (and their superpartners), it will be important to
include their order $M_{\rm GUT}$ masses.  We use the method of mass insertions and
then sum over loops with all number of insertions.  For simplicity, we first present 
the calculation for a gauged U(1) broken at the GUT scale and later generalize to 
minimal SU(5) by summarizing the effects of the additional group structure.

\FIGURE[t]{
\epsfxsize=4.5in
\centerline{\epsfbox{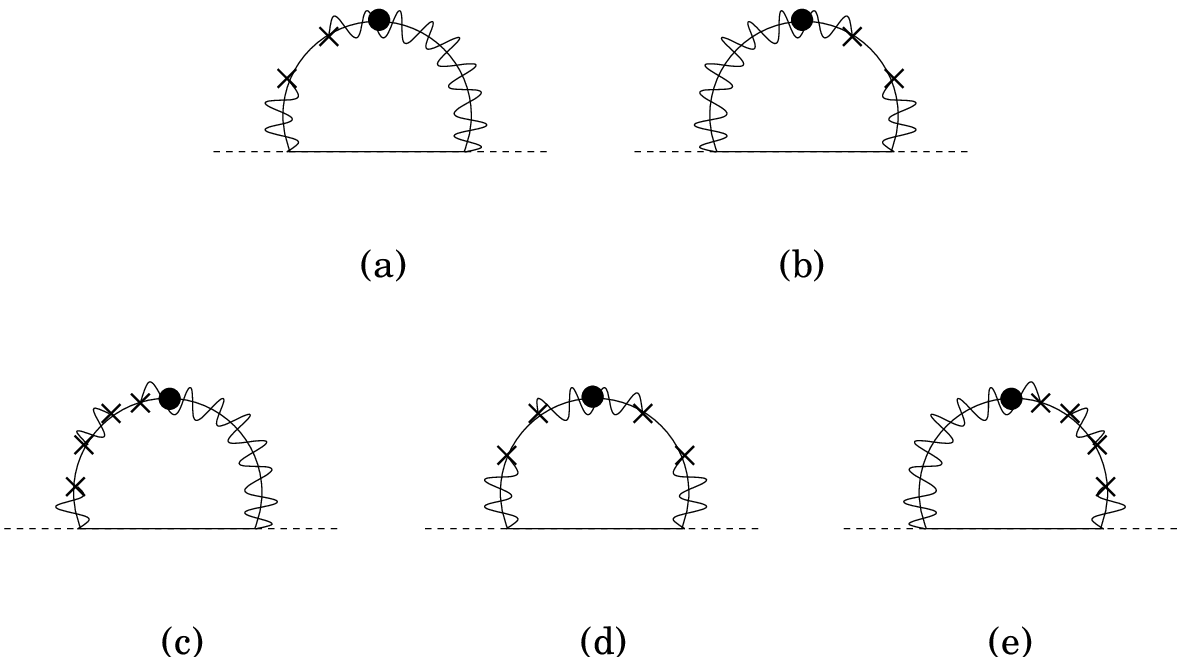}}
\caption{Loop contributions to squared scalar masses with (a,b) one and (c,d,e) two
insertions of $M_{\rm GUT}^2\times$(the adjoint propagator).}
\label{GUTloop}}

We assume a field $\phi$ of charge $+1$ living on the matter boundary has a scalar vev
$\langle\phi\rangle=M_{\rm GUT}/(\sqrt{2}g)$.  To calculate the loop contribution 
to light scalar masses due to the U(1) vector multiplet and the operators 
(\ref{opsb})-(\ref{opsf}),
we must calculate diagrams like those in Fig. \ref{loop}.  In order to include the 
effects of the GUT masses, we include mass insertions on the gaugino propagators as in
Fig. \ref{GUTloop} where the crosses represent insertions of $M_{\rm GUT}$ and the
solid lines in the upper part of the loop are propagators of $\psi$, the fermionic
partner of $\phi$.  The diagram in Fig. \ref{GUTloop}(a) has the functional form
(at zero external momentum)
\begin{eqnarray}
\int \frac{d^4 q}{(2\pi)^4} 
	&& {\rm tr} \{ \sqrt{2} g_{(5)} P_L \left[ \mathcal{P}(q;0,0)
	\sqrt{2} g_{(5)} \frac{M_{\rm GUT}}{\sqrt{2}g} \frac{-1}{\qslash} 
	\sqrt{2} g_{(5)} \frac{M_{\rm GUT}}{\sqrt{2}g} \right]  \nonumber\\\nonumber\\
	&&\times \;\mathcal{P}(q;0,L) \frac{F^2}{M_*^5} \qslash \mathcal{P}(q;L,0) 
	\sqrt{2} g_{(5)} \frac{-1}{\qslash} \}
\label{GUTinsert}
\end{eqnarray}
where $g_{(5)}$ and $g$ are the five- and four-dimensional gauge couplings 
respectively, the quantity in square brackets represents the insertion on the gaugino 
propagator and the integral is done in Euclidean space.  For the gaugino, 
we use the mixed position/momentum space propagator which 
appeared in \cite{KKS}:
\begin{eqnarray}
\mathcal{P}(q;a,b)=&& \frac{2}{L} \sum_{m=0}^{\infty} 
\left[P_L {1 \over \sqrt{2}^{\delta_{n0}}}\cos(\frac{m\pi}{L} a)
	- P_R \sin(\frac{m\pi}{L} a)\right]
{\qslash-i\gamma_5 \frac{m\pi}{L} \over q^2+(\frac{m\pi}{L})^2}\nonumber\\\nonumber\\
&&\times\;\left[P_R {1 \over \sqrt{2}^{\delta_{m0}}}\cos(\frac{m\pi}{L} b) 
	+ P_L \sin(\frac{m\pi}{L} b) \right] \ .
\end{eqnarray}
Since $\mathcal{P}(q;0,0)$ commutes with $1/\!\!\qslash$, we see that Eq. (\ref{GUTinsert})
also represents Fig. \ref{GUTloop}(b).

\FIGURE[t]{
\epsfxsize=3.5in
\centerline{\epsfbox{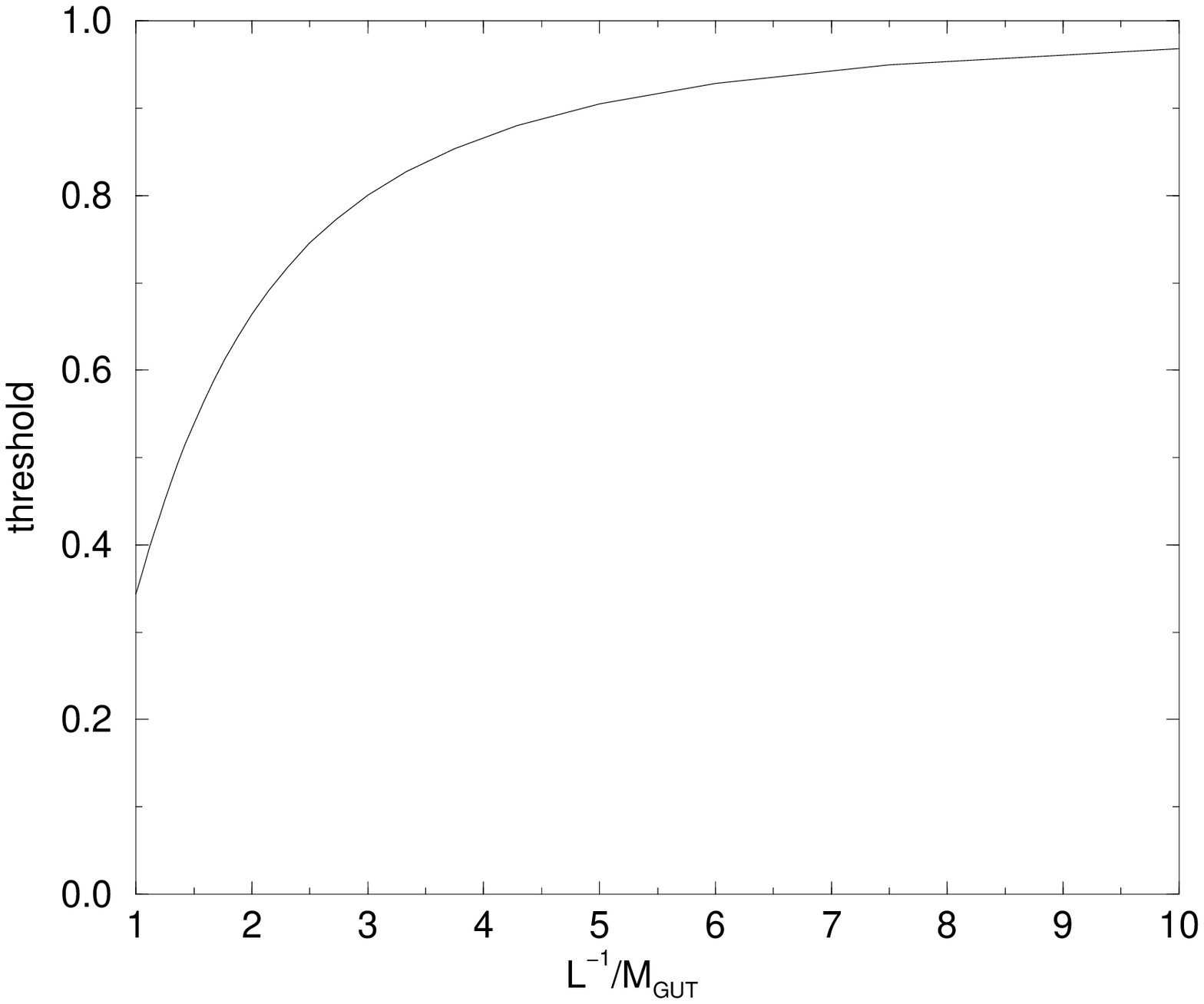}}
\caption{The squared scalar mass contribution from a GUT-scale gauge multiplet.  The
curve is normalized to the same contribution with $M_{\rm GUT}=0$.}
\label{GUTplot}}

Now we can generalize to any number of insertions.  There are $n$ diagrams 
with $(n-1)$ insertions that have the form
\begin{equation}
\int \frac{d^4 q}{(2\pi)^4} {\rm tr}\left\{ 2 g_{(5)}^2 P_L 
        \mathcal{P}(q;0,L) \frac{F^2}{M_*^5} \qslash \mathcal{P}(q;L,0) 
        \frac{-1}{\qslash} 
	\left[ \mathcal{P}(q;0,0)
        M_{\rm GUT}^2 L \frac{-1}{\qslash} 
        \right]^{(n-1)}
	\right\}
\label{GUTn-insert}
\end{equation}
where we used the relation $g_{(5)}^2=g^2/L$.  Using
\begin{equation}
\sum_{n=1}^{\infty} n x^{n-1} = \frac{\partial}{\partial x} \sum_{n=1}^{\infty} x^n 
	= \frac{\partial}{\partial x} \frac{1}{1-x} = \frac{1}{(1-x)^2}
\end{equation}
we sum over all GUT insertions and get
\begin{eqnarray}
&& \int \frac{d^4 q}{(2\pi)^4} {\rm tr}\left\{ 2 g_{(5)}^2 P_L
        \mathcal{P}(q;0,L) \frac{F^2}{M_*^5} \qslash \mathcal{P}(q;L,0)
        \frac{-1}{\qslash}
        \frac{1}{\left[ 1 + M_{\rm GUT}^2 L\mathcal{P}(q;0,0) \frac{-1}{\qslash}
        \right]^2}
        \right\}\nonumber\\\nonumber\\
&=& \int \frac{d^4 q}{(2\pi)^4} 4 g^2 \frac{F^2}{M_*^5 L} q^2
	\left(\sum_{m=0}^{\infty} \frac{1}{2^{\delta_{m0}}}\frac{(-1)^m}{q^2 + (m\pi/L)^2}
	\right)^2
	\frac{1}{\left[ 1 - 2 M_{\rm GUT}^2 \sum_{m=0}^{\infty} 
	\frac{1}{2^{\delta_{m0}}}\frac{1}{q^2 + (m\pi/L)^2}
        \right]^2} \nonumber\\\nonumber\\
&=& \frac{g^2}{2\pi^2} \frac{F^2}{M_*^5 L^3} 
	\int d\tilde{q} \frac{\tilde{q}^3}{\sinh^2{\tilde{q}}}
	\frac{1}{\left[ 1 - (M_{\rm GUT} L)^2 (\coth{\tilde{q}})/\tilde{q}
        \right]^2}
\end{eqnarray}
which can be computed numerically.  In Fig. \ref{GUTplot} we have plotted the ratio
of this integral to itself with $M_{\rm GUT}=0$ over a range of $(M_{\rm GUT} L)^{-1}$.
We see that the non-zero masses of GUT fields should not be ignored for compactification
scales up to a factor of a few times the GUT scale.

It is relatively straight forward to generalize the above result to a minimal SU(5) GUT.
Take $\phi$ to now be in the adjoint representation with a vev
\begin{equation}
\langle\phi\rangle = v_{\rm GUT} 
   \left(
	\begin{array}{ccccc}
	1 & \phantom{0} & \phantom{0} &\phantom{0} &\phantom{0} \\
	\phantom{0} & 1 & \phantom{0} &\phantom{0} &\phantom{0} \\
	\phantom{0} &\phantom{0} & 1 & \phantom{0} &\phantom{0} \\
	\phantom{-0} & \phantom{-0} &\phantom{-0} & - \frac{3}{2} &\phantom{0} \\
	\phantom{0} & \phantom{0} &\phantom{0} & \phantom{0} & - \frac{3}{2} \\
	\end{array}
   \right)
= \sqrt{\frac{6}{5}}\frac{M_{\rm GUT}}{g}t^{24}
\end{equation}
where $t^{24}$ is a generator of SU(5) normalized such that 
tr$\left[t^{24}t^{24}\right]=\frac{1}{2}$.  Now the two quark-squark-gaugino 
vertices contribute the factor $2 g^2 (t^a t^b)^i_j$.  The insertions of
$M_{\rm GUT}$ now come with structure constants $f^{24 a c}$.  In the normalization
we chose, $\sum_c (f^{24 a c}f^{24 b c})= \sqrt{5/12}\,\delta'_{ab}$, where $\delta'=1$ 
when $a=b=x$ with $x$ representing the broken generators 
SU(5)/[SU(3)$\times$SU(2)$\times$U(1)], and $\delta'=0$ otherwise.  This change gives
us the same $M_{\rm GUT}/\sqrt{2}g$ insertions as before.  The only difference
from the Abelian case is a ``partial Casimir'' which depends on the representation
(see Table \ref{Casimir-table} in Sec. \ref{pheno}) multiplying the integral.

In the calculation done above, all additional GUT fields or interactions have
been ignored.  For instance, in order to get a vev, $\phi$ must appear in superpotential
terms, the simplest being a Majorana mass term.  These new interactions can have an effect
the threshold calculation.  The most significant of which could 
come from the mechanism
which splits the Higgs doublets from the rest of their multiplet 
(``doublet-triplet splitting'').
The new multiplets required for such a mechanism could also contribute to Higgs soft masses
at two loops if the new representations are large enough.  Finally, other GUT groups will 
obviously have different contributions.  For instance, at $r=2$, the gaugino-assisted
contribution in minimal SO(10) would be approximately universal for all 
chiral matter.  The group $E_6$ could potentially have a universal 
contribution to {\it all} matter fields.  However,
realistically we expect things like doublet-triplet splitting and 
contributions from $D$ terms to destroy this naive degeneracy.


\section*{Acknowledgments}

We are grateful to Zackaria Chacko, Hsin-Chia Cheng, Markus Luty,
Erich Poppitz, Martin Schmaltz, Yael Shadmi, and Carlos Wagner
for many useful discussions.  In addition, we thank Hsin-Chia Cheng and 
Tim Tait for comments on the final draft.  We also thank Zackaria Chacko and 
Markus Luty for communicating some of the results of their paper 
Ref.~\cite{RMSB} 
to us before it was completed.  GDK would like to thank the Argonne high energy 
theory group and the organizers of the Argonne Theory Institute 2000 for their 
hospitality.  We also thank the Aspen Center for Physics where this work was 
completed.  DEK is supported in part by the DOE under contracts 
DE-FG02-90ER40560 and W-31-109-ENG-38.  GDK is supported in part by the 
DOE under contracts DOE-ER-40682-143 and DE-FG02-95ER40896.

\end{document}